\documentclass[
    ,final            % use final for the camera ready runs
  ]
  {aipproc}

\layoutstyle{8x11single}

\newcommand{\eV}{\mbox{eV}}
\newcommand{\keV}{\mbox{keV}}

\newcommand{\GeV}{\mbox{GeV}}
\newcommand{\TeV}{\mbox{TeV}}
\newcommand{\cu}{\mbox{c.u.}}

\newcommand{\magic}{\mbox{MAGIC}}

\newcommand{\asm}{\mbox{{\it RXTE}/ASM}}

\newcommand{\vhe}{\mbox{VHE}}
\newcommand{\gray}{\mbox{$\gamma$-ray}}

\newcommand{\xrays}{\mbox{X-rays}}

\newcommand{\xray}{\mbox{X-ray}}
\newcommand{\Fvar}{\mbox{$F_{var}$}}

\newcommand{\etal}{\mbox{et al}}

\begin{document}

\title{Study of the Flux and Spectral Variations in the VHE Emission from the Blazar Markarian 501, with the MAGIC Telescope}

 %               \texttt{http://www.aip..org/pacs/index.html}>}

\author{D.Paneque}{
  address={Stanford Linear Accelerator Center (SLAC) / Max-Planck-Institut f\"{u}r Physik, M\"{u}nchen (MPI)}
}
\author{on behalf of the MAGIC collaboration}{
  address={updated author list can be found at \url{http://wwwmagic.mppmu.mpg.de/collaboration/members}}
}

\begin{abstract}
The blazar Markarian 501 (Mrk 501) was observed above 100 \GeV\ with the MAGIC Telescope during 
May, June and July 2005. 
The high sensitivity of the instrument made possible the  detection of the source with high significance 
in each of the observing nights.
During this observational campaign, 
the emitted gamma-ray flux from Mkn 501 was found to vary by one order of magnitude, 
and showed a high correlation with spectral changes. 
Intra-night flux variability was also observed, with flux-doubling times of $\sim$ 2 minutes. 
 The data showed a clear
evidence of a spectral peak (in the $\nu F \nu$ representation) 
during the nights when the gamma-ray activity was highest. The location of this spectral feature was 
found to be correlated with the emitted gamma-ray flux. 
In these proceedings we discuss some of the results of this unprecedented spectral and temporal analysis 
of Mrk 501 observations in the very high energy range.

\end{abstract}

\maketitle

%%%%%%%%%%%%%%%%%%%%%%%%%%%%%%%%%%%%%%%%%%%%
%% MAINMATTER
%%%%%%%%%%%%%%%%%%%%%%%%%%%%%%%%%%%%%%%%%%%%

\section{Observation and Results}

The observations of Mrk 501 in the Very High Energy (\vhe) domain were carried out 
with the Major Atmospheric Gamma-ray Imaging Cherenkov ({\bf \magic}) Telescope (see \cite{Baixeras2004,PanequeThesis,Cortina2005,GaugThesis} for details of 
MAGIC). The net observation time on the source is 31.6 hours (24 nights) between May and 
July 2005. Quasi-simultaneously to the \magic\ data, Mrk 501 was regularly observed 
with KVA as a part of Tuorla Observatory's blazar 
monitoring program\footnote{\tt http://users.utu.fi/kani/1m/.}. 
And we also used data taken with the 
{\it RXTE} satellite's All-Sky-Monitor (\asm)\footnote{
The data are publicly available at \mbox{\tt http://heasarc.gsfc.nasa.gov/xte\_weather/}.}.

\begin{figure}
  \includegraphics[height=.65\textheight]{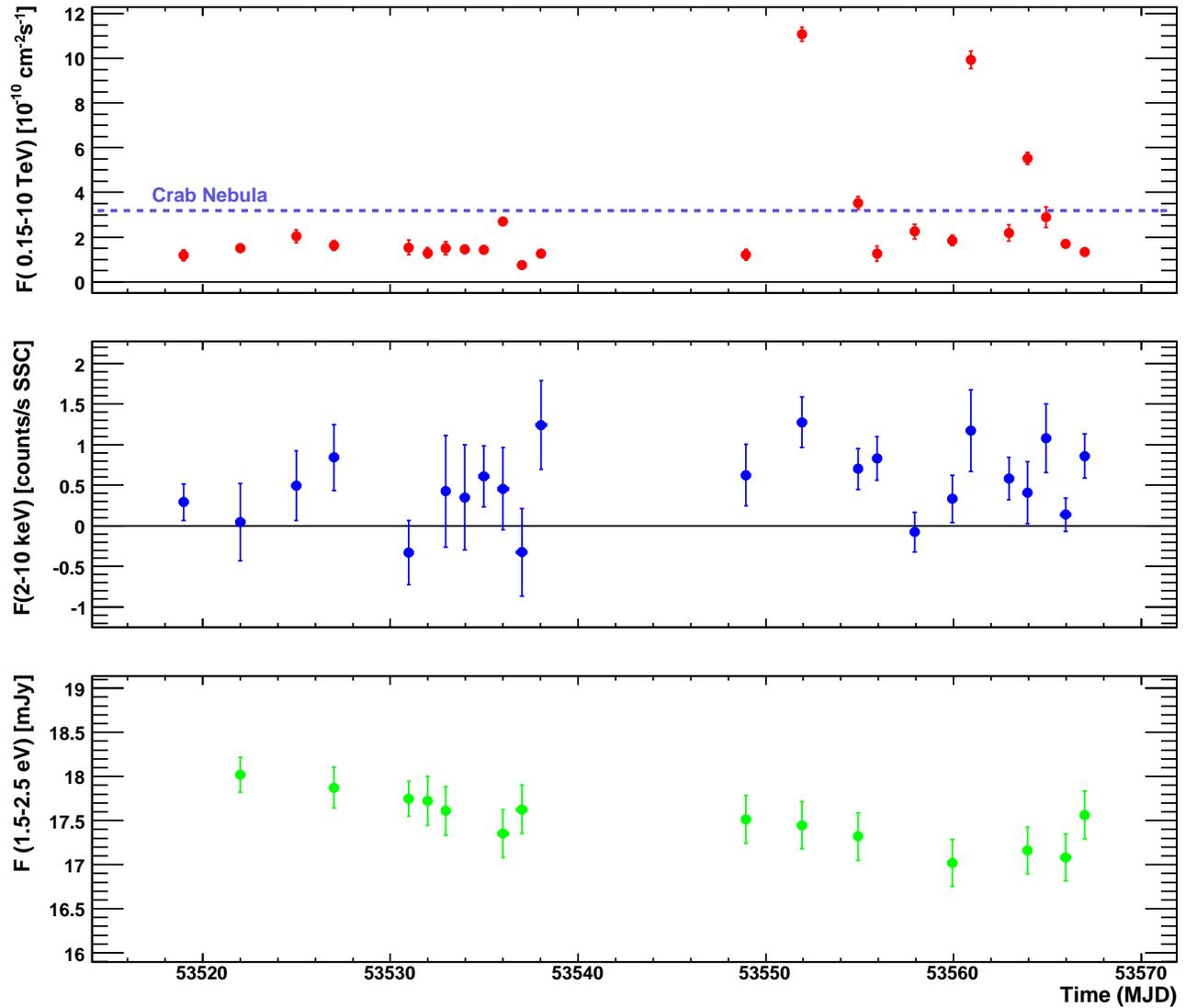}
\caption{Multi-frequency LC during the MAGIC observations of 
\mbox{Mrk 501} (May-July 2005). {\it Top)} MAGIC flux above 0.15 TeV. 
The Crab flux is also shown for comparison (lilac dashed horizontal line). 
{\it Middle)} $RXTE$/ASM 2-10 \keV\ flux.  
{\it Bottom)} KVA $\sim$1.5-2.5 \eV\ flux. 
Error bars denote 1$\sigma$ statistical uncertainties. The X-ray/optical data were selected to match 
the MAGIC data within a time window of 0.2 days.}
\label{Fig1}
\end{figure}

{\bf The overall Light Curve} (LC) of \mbox{Mrk 501} during the \magic\ observation campaign is shown in 
Fig. \ref{Fig1}. 
The observed flux is shown in three energy bands: \vhe\ 
(0.15\GeV-10\TeV), \xrays\ (2\keV-10\keV), and optical (1.5\eV-2.5\eV) as 
measured by \magic, \asm\, and KVA, respectively. The \xray\ and optical fluxes are computed 
as weighted averages using \asm\ and KVA measurements taken simultaneously with the \magic\ 
observations plus/minus a time tolerance  of 0.2 days. A smaller time tolerance substantially 
decreases the number of \xray\ points that can be used. The \gray\ flux level of the Crab Nebula 
(lilac-dashed horizontal line in the top plot) is also shown in the upper plot of Fig. \ref{Fig1} 
for comparison. The Crab nebula flux was obtained by applying the very same analysis used 
for Mrk 501 to the MAGIC Crab nebula data taken during December 
2005 under observing conditions similar to those for Mrk 501. The estimated Crab Nebula flux 
level is therefore roughly affected by the same systematics as the fluxes obtained for Mrk 501. 
We found $F_{\rm Crab}$($>$$0.15\,$TeV)=(3.2$\pm$0.1)$\times$$10^{-10}$ cm$^{-2}$s$^{-1}$, 
thereafter referred to as Crab Unit (\cu). Fig. \ref{Fig1} 
shows that the \vhe\ flux from 
\mbox{Mrk 501} was about 0.5 \cu\ during most of the observation 
nights; yet there are significant deviations from this mean \gray\ emission, 
with large flux variations occurring in consecutive nights.

\begin{figure}
  \includegraphics[height=.35\textheight]{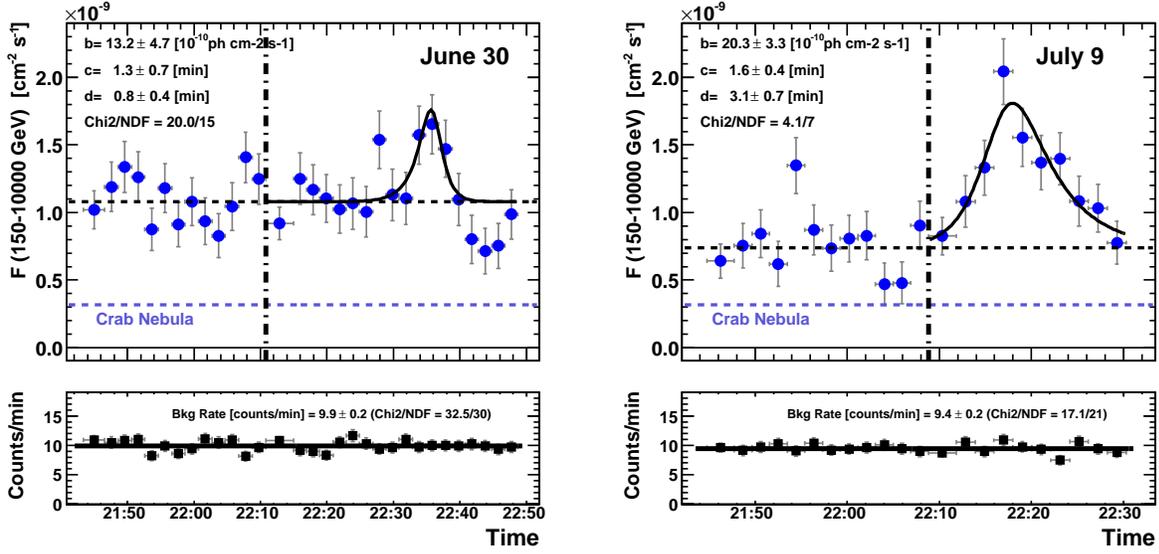}
  \caption{Integrated-flux LCs of \mbox{Mrk 501} for the flare nights of June 30 and July 9. 
Horizontal bars represent the 2-minute time bins, and vertical bars denote 1$\sigma$ statistical 
uncertainties. For comparison, the Crab emission is also shown as a lilac dashed horizontal line. 
The vertical dot-dashed line divides the data into 'stable' (i.e., pre-burst) and 'variable' (i.e., 
in-burst) emission emission. The horizontal black dashed line represents the average of the 'stable' 
emission. The solid black curve represents the best-fit flare model (see eq. \ref{eq_fitflare}).
The insets report the fit parameters and goodness of the fit.
The bottom plots show the mean background rate during each of the 2-minute bins of the LCs.
The insets report the mean background rate during the entire night, resulting from a 
constant fit to the data points. The goodness of such fit is also given.
The background rates are constant along the entire night. Consequently, the variations 
seen in the upper panels of the middle and right-hand plot correspond to actual 
variations of the \vhe\ \gray\ flux from Mrk501, thus 
ruling out detector instabilities and/or atmospheric changes.}
\label{Fig2}
\end{figure}

During the two nights with the highest \vhe\ activity ($>$3\cu), namely June 30 and July 9, Mrk 501 
clearly showed {\bf intra-night flux variations}. The corresponding LC in the 0.15-10 TeV band is 
shown in Fig. \ref{Fig2} with a time 
binning of $\sim$2 minutes. A constant line fit to the whole LC 
gives a $\chi^2/NDF=47.9/30$ (probability $p=2.0 \times 
10^{-2}$) for the night of June 30, and  a $\chi^2/NDF=80.6/21$ ($p=6.4\cdot10^{-9}$) for the 
night of July 9. Therefore, the emission above 150 \GeV\ during the two nights is statistically 
inconsistent with being constant. The burst's amplitude and duration, 
as well as its rise/fall times, were quantified according to the following function \cite{thomasthesis}:

\vspace{0.5cm}
\begin{equation}
\label{eq_fitflare}
F(t) = a + \frac{b}{2^{-\frac{t-t_0}{c}} + 2^{\frac{t-t_0}{d}}}
\end{equation} 
\vspace{0.5cm}

This model parametrizes a flux variation (flare) superposed on a stable emission: $F(t)$ 
asymptotically tends to $a$ when $t \rightarrow \pm \infty$. The parameter $a$ is the assumed 
constant flux at the time of the flare (cf. the horizontal black dashed lines in Fig. \ref{Fig2}); 
$t_0$ is set to the time 
corresponding to the point with the highest value in 
the LC; and $b,c,d$ are left free to vary. The latter two parameters denote the flux-doubling 
rise and fall times, respectively. The resulting fit parameters are reported in the insets 
of Fig. \ref{Fig2}, {\bf showing flux-doubling times of the 
order of 2 minutes}. This is the fastest flux variability ever observed from Mrk 501. 
These bursts were also studied in four non-overlapping energy ranges; 0.15-0.25 \TeV, 
0.25-0.6 \TeV, 0.6-1.2 \TeV\ and $>$1.2 \TeV. The main outcome of this study is 
\textit{a)} the burst from June 30 is significant only in the energy range 0.25-0.6 \TeV, while 
the burst from July 9 is significantly observed in all energy ranges; 
\textit{b)} in the July 9 burst, the highest energies are 
delayed by 4 $\pm$ 1 minutes with respect to the lowest energies; 
and \textit{c)} in the July 9 burst, 
the relative amplitude of the flux variation increases with energy. Further details 
from these studies can be found elsewhere \cite{Mrk501Publication}.

\begin{figure}
  \includegraphics[height=.31\textheight]{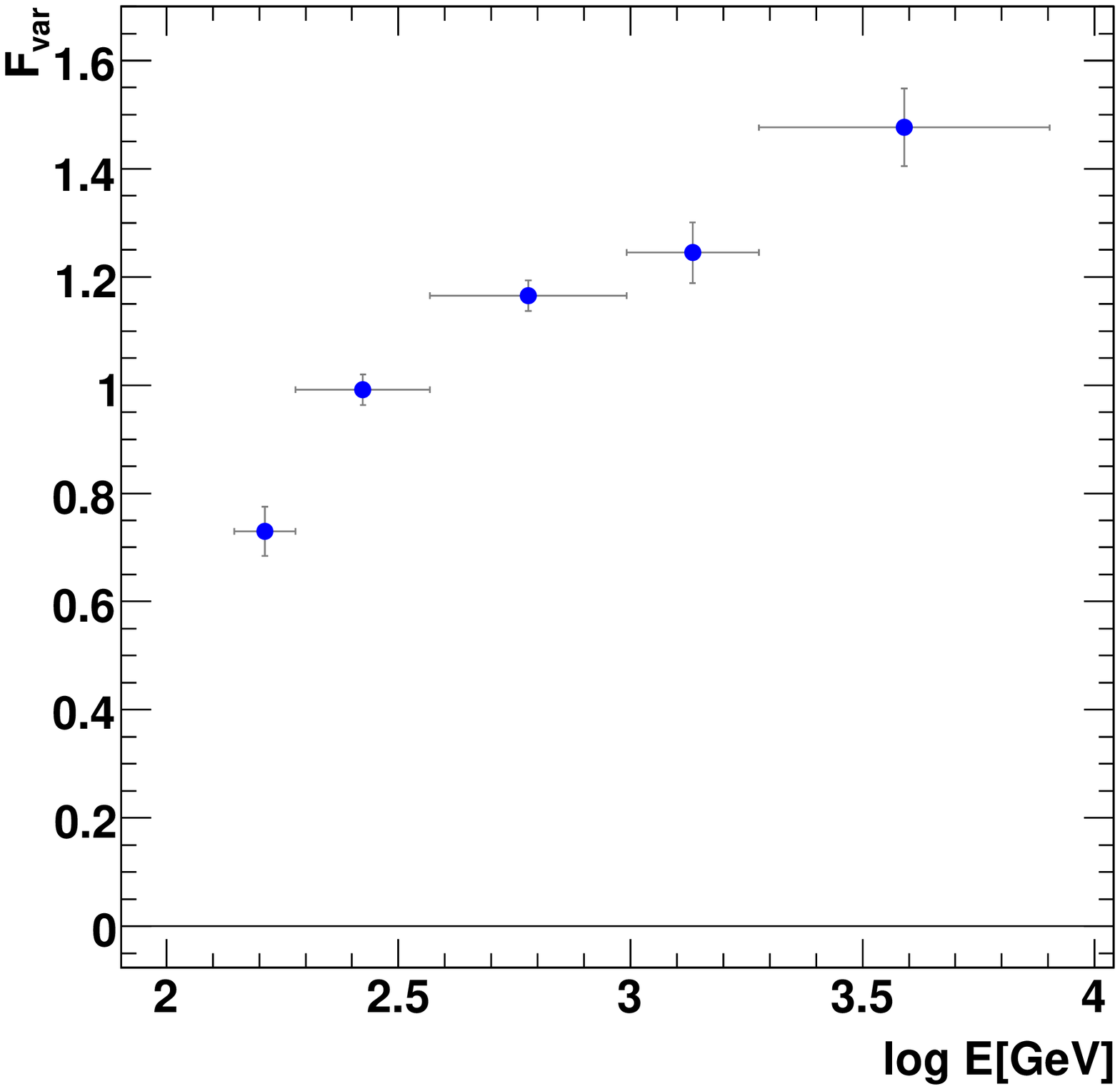}
  \includegraphics[height=.31\textheight]{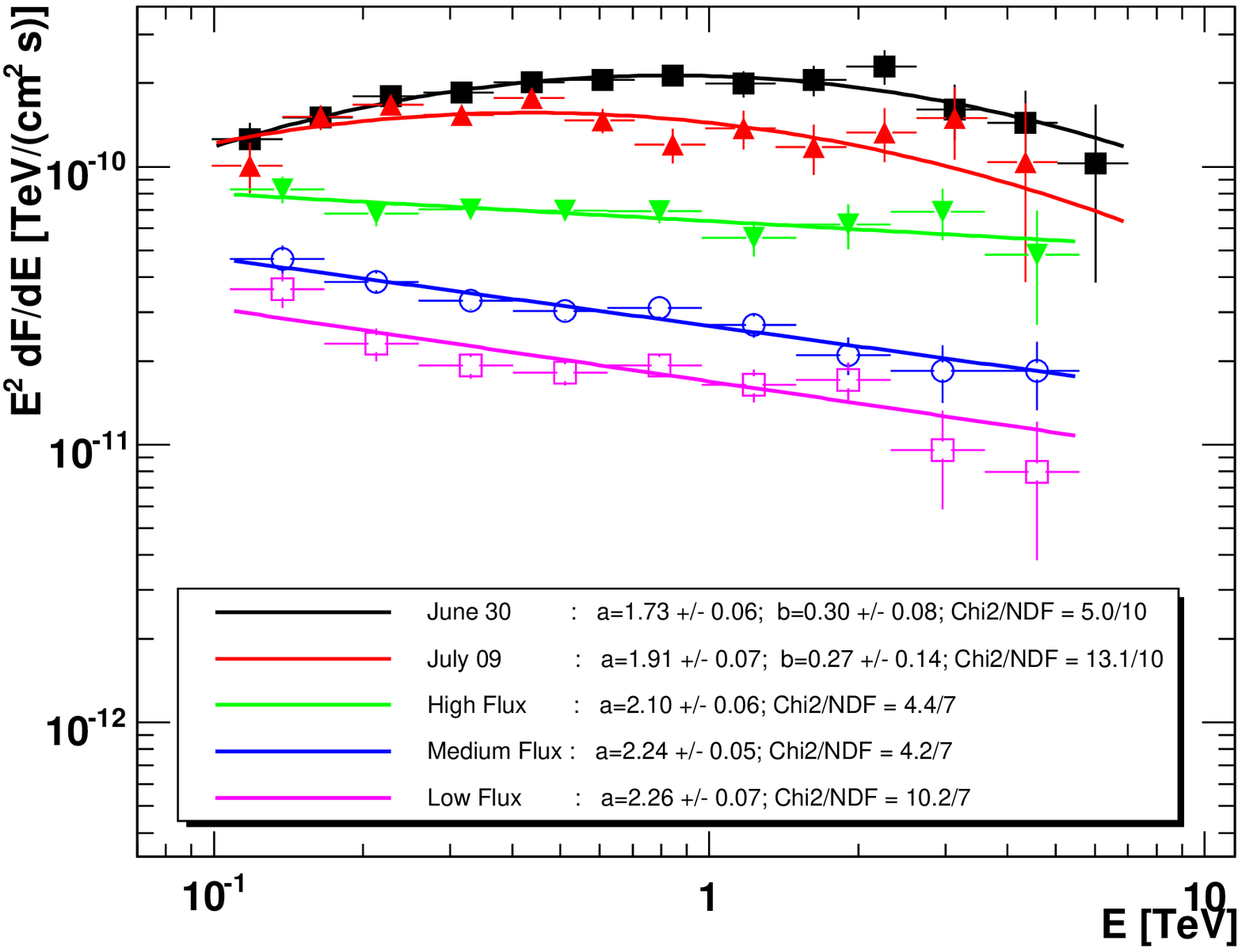}
  \caption{\texttt{Left-hand}: Fractional variability parameter, calculated as described in \cite{Vaughan2003}. 
Vertical bars denote 1$\sigma$ uncertainties, horizontal bars indicate the width of each energy bin.
\texttt{Right-hand}: Energy spectra of \mbox{Mrk 501} for the flaring nights of 
June 30 (black squares) and July 9 (red up-triangles), and for three data sets which group nights  
according to whether their integral flux above 150 \GeV, $F_{150\,{\rm GeV}}$ (measured in {\it Crab Units} (\cu)), 
was high ($1.0~\cu$$<$$F_{150\,{\rm GeV}}$; green down triangles),
medium ($0.5~\cu$$<$$F_{150\,{\rm GeV}}$$< 1.0~\cu$; blue open circles), or low 
($F_{150\,{\rm GeV}}$$<$$0.5~\cu$; pink open squares).
Vertical bars denote 1$\sigma$ 
uncertainties, horizontal bars denote energy bins. Lines show best fits using log-parabolic (for flare nights) 
and power-law (for high/medium/low flux levels) functions. The insets report the spectral indices derived 
from the fit, as well as the goodness of such fit. 
The spectra are corrected for EBL extinction using \citep{EBLKneiske}'s 'Low' EBL model.} 
\label{Fig3}
\end{figure}

Mrk 501 showed {\bf energy-dependent flux variations} throughout the entire \magic\ observational campaign. 
We followed the prescription given in \cite{Vaughan2003} to quantify the flux variability by means of 
the fractional variability parameter \Fvar, as a function of energy. The outcome of this calculation 
is shown in the left-hand plot of Fig. \ref{Fig3}; the flux variability increases with energy.
The same tendency was reported at \xray\ frequencies, but with lower values of \Fvar\ \cite{gliozzi2006}.

During our observations, the \vhe\ emission of \mbox{Mrk 501} was very dynamic, 
showing significant {\bf spectral variability} on a timescale of days. 
Nevertheless, most of the data are well described by a simple Power-law (PL) function. 
This does not hold for the two flaring nights of June 30 and 
July 9 which show clear spectral curvature and thus they are fit with a log-parabolic (LP) function.

\vspace{0.5cm}
\begin{equation}
\label{fitfuncs}
PL: ~ {dF \over dE} = K_0 \cdot \Big(\frac{E}{0.3 \TeV}\Big)^{-a} 
~~~~~~~~~~~~~~~~~~~~~~~~~~~~~~~~ 
LP: ~ {dF \over dE} = K_0 \cdot \Big(\frac{E}{0.3 \,TeV}\Big)^{-a 
           - b \cdot log_{10} \big(\frac{E}{0.3\, TeV}\big)}
\end{equation}
\vspace{0.5cm}

Here, $K_0$ is a normalization factor, $a$ is the spectral index at 0.3 TeV, and $b$ is 
a curvature parameter.

The right-hand plot of Fig. \ref{Fig3} shows the {\bf energy spectra} (EBL corrected) of Mrk 501 for 
the flaring nights (June 30 and July 9) for three different flux levels, {\it low}, 
{\it medium} and {\it high}, 
which contain  12, 8 and 2 (``non-flaring'') nights respectively. See caption of Fig. \ref{Fig3} 
for definition of flux levels. The data shows very clearly a hardening of the spectra with 
increasing flux. Besides, the data from the flaring nights suggest the presence of a {\bf spectral peak}. 
The location of this peak, determined using the fit parameters and errors from the log-parabolic fit, 
is 0.85 $\pm$ 0.13 \TeV\ and 0.44 $\pm$ 0.08 \TeV\ for June 30 and July 9, respectively.
If this peak exists for the data sets {\it low}, 
{\it medium} and {\it high}, then its location is certainly below 0.1 \TeV. Therefore, the location 
of such spectral peak is very probably correlated with the source luminosity.

%\begin{figure}
%  \includegraphics[height=.20\textheight]{Figures/LC_2flares_E150_2min_bkgrates_v3_FitInfo.eps}
%  \caption{Integrated-flux LCs of \mbox{Mrk 501} for the flare nights of June 30 and July 9. 
%Horizontal bars represent the 2-minute time bins, and vertical bars denote 1$\sigma$ statistical 
%uncertainties. For comparison, the Crab emission is also shown as a lilac dashed horizontal line. 
%The vertical dot-dashed line divides the data into 'stable' (i.e., pre-burst) and 'variable' (i.e., 
%in-burst) emission emission. The horizontal black dashed line represents the average of the 'stable' 
%emission. The solid black curve represents the best-fit flare model (see eq. \ref{eq_fitflare}).
%The insets report the fit parameters and goodness of the fit.
%The bottom plots show the mean background rate during each of the 2-minute bins of the LCs.
%The insets report the mean background rate during the entire night, resulting from a 
%constant fit to the data points. The goodness of such fit is also given.}
%\label{Fig2}
%\end{figure}

\section{Conclusions}

We have undertaken a systematic study of the temporal and spectral variability 
of the nearby blazar \mbox{Mrk 501} with the MAGIC telescope at energies $>$ 0.1 \TeV.
%The sensitivity of MAGIC allowed us to carry out these studies with unprecedented precision.
During 24 observing nights between May and July 2005, 
all of which yielded significant detections, we measured 
fluxes and spectra at levels of baseline activity ranging from $<$0.5 
to $>$1 \cu. During two nights, on June 30 and July 9, \mbox{Mrk 501} 
underwent into a very active state with a $\gamma$-ray emission $>$3 \cu, 
and flux-doubling times of $\sim$2 minutes. 
%The $\sim$20-minute long flare of 
%July 9 showed an indication of a 
%4$\pm$1 min time delay between the peaks of F($<$0.25 TeV) and F($>$1.2 TeV), which
%may indicate a progressive acceleration of electrons in the emitting plasma blob. 
An overall trend of harder 
spectra for higher flux was clearly seen.
The VHE $\gamma$-ray variability was found to increase with energy, 
and it is significantly higher 
than the variability at X-ray frequencies. 
A spectral peak, at a location dependent on source luminosity, was clearly observed during the active 
states. 
All these features are 
naturally expected in synchro-self-Compton (SSC) models of blazar VHE emission.
See \cite{Mrk501Publication} for further details from these observations.

%%%%%%%%%%%%%%%%%%%%%%%%%%%%%%%%%%%%%%%%%%%%%%%%
%% BACKMATTER
%%%%%%%%%%%%%%%%%%%%%%%%%%%%%%%%%%%%%%%%%%%%%%%%

%\begin{theacknowledgments}
 
%\end{theacknowledgments}

%%%%%%%%%%%%%%%%%%%%%%%%%%%%%%%%%%%%%%%%%%%
%% The following lines show an example how to produce a bibliography
%% without the help of the BibTeX program. This could be used instead
%% of the above.
%%%%%%%%%%%%%%%%%%%%%%%%%%%%%%%%%%%%%%%%%%%

%%\cleardoublepage

\end{document}